\def\order#1{{\cal O}\left(#1\right)}
\def\ba{\begin{eqnarray}}
\def\ea{\end{eqnarray}}
\def\eff{\mbox{eff}}

\def\pPs{\mbox{p-Ps}}
\def\oPs{\mbox{o-Ps}}
\def\oDm{\mbox{o-Dm}}
\def\oMu{\mbox{o-Mu}}

\documentstyle[12pt,psfig]{article}

\begin{document}

\begin{flushright}
BNL-HET-99/37\\
hep-ph/9911410\\
November 1999
\end{flushright}

\begin{center}
{\Large \bf Decays of Positronium\footnote{Talk given at the 14th
Intl.~Workshop on High Energy Physics and Quantum Field Theory 
(QFTHEP99), May 1999, Moscow, Russia.}}\\

\vspace{4mm}

Andrzej Czarnecki$^a$ and Savely G. Karshenboim$^b$
\\[2mm]
$^a${\em Physics Department,
Brookhaven National Laboratory\\  Upton, New York 11973, USA}
\\[2mm]
$^b${\em D. I. Mendeleev Institute for Metrology, St. Petersburg,
Russia}
\\
\end{center}

%                       Abstract
\begin{abstract}
We briefly review the theoretical and experimental results concerning
decays of positronium.  Possible solutions of the ``orthopositronium
lifetime puzzle'' are discussed.  Positronium annihilation into
neutrinos is examined and disagreement is found with previously
published results.  
\end{abstract}

\section{Introduction}
Positronium (Ps), an electron-positron bound state, is the lightest
known atom.  Its gross spectrum is similar to that of the hydrogen,
except for a different reduced mass\footnote{We use $\hbar=c=1$,
$\alpha=e^2/4\pi\simeq 1/137$, and denote electron's mass and charge
by $m_e$ and $-e$, respectively.} $m_{\sc R}=m_e/2$.  Ps is of
significant theoretical and experimental interest.  In contrast to the
hydrogen, it does not contain a proton and its theoretical description
is not limited by hadronic uncertainties, at least at the present
level of experimental accuracy.  Spectrum and lifetimes of Ps states
can be predicted within Quantum Electrodynamics (QED) with very high
accuracy.  Recently, significant progress in QED theory of positronium
spectrum has been achieved.  Effects $\order{\alpha^6m_e}$
\cite{Adkins97,Hoang:1997ki,PhK,Czarnecki:1998zv,Czarnecki:1999mw} and
$\order{\alpha^7\ln^2(\alpha)m_e}$ \cite{DL,PK99,MY} were
calculated.  Some progress has also been made in the study of the
positronium annihilation rate, which we will discuss below.  Combined
with precise experimental measurements, these results provide unique
opportunities for testing bound-state theory based on QED.  For
example, one can study recoil effects in detail because in Ps they are
not suppressed by a small mass ratio (as opposed to e.g. muonium).

In addition, precise studies of positronium decays provide bounds on
exotic particles, such as axions, paraphotons, or millicharged
particles.  This short note is devoted to a review of the standard and
some exotic positronium decay channels.  In particular, we will be
interested in the decay into neutrinos, where we find disagreement
with previously published results.

\section{Photonic decays of positronium}
\subsection{Parapositronium decays}

\begin{figure}[htb] 
\hspace*{33mm}
\begin{minipage}{16.cm}
\vspace*{3mm}
\begin{tabular}{cc}
\psfig{figure=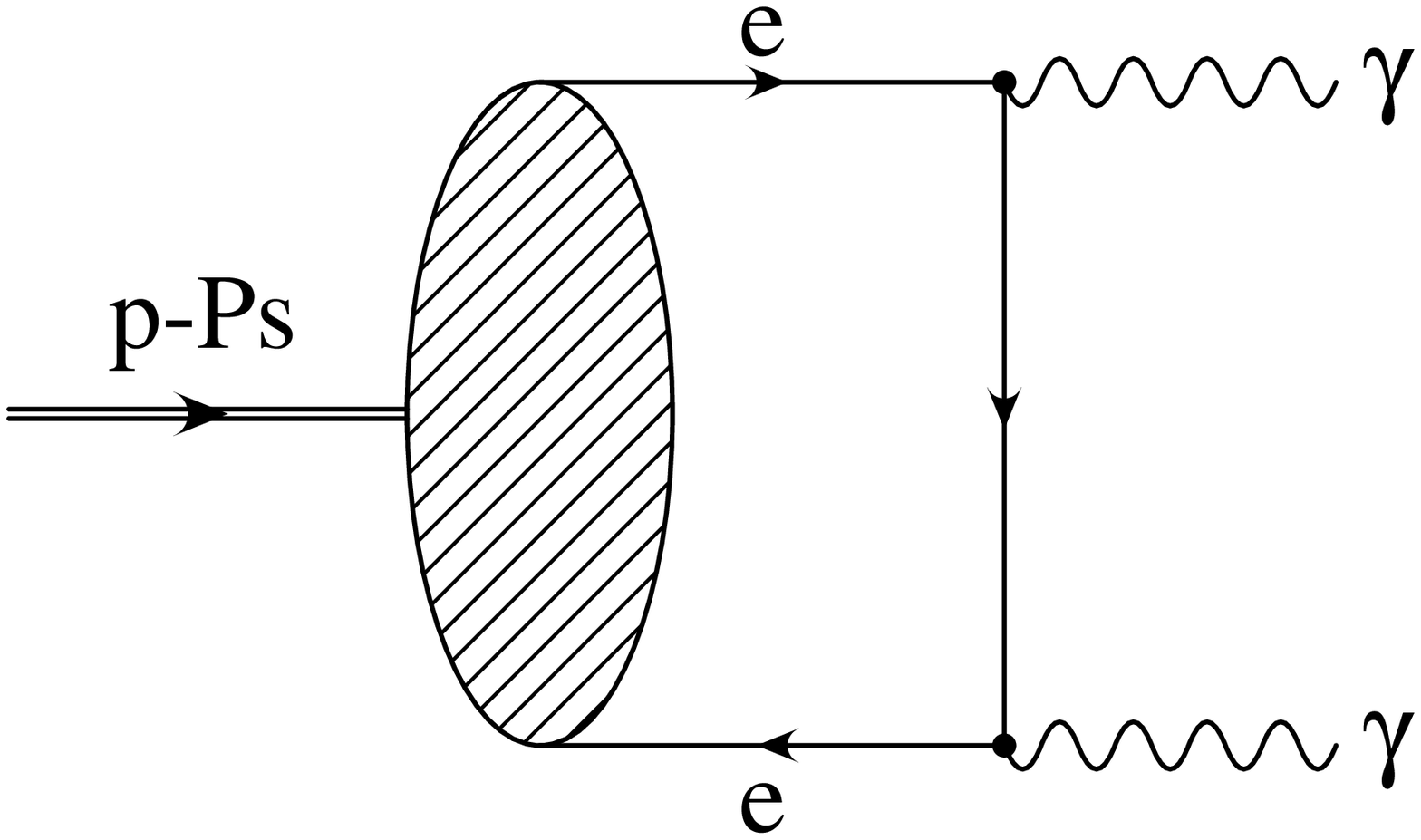,width=35mm,bbllx=72pt,bblly=291pt,%
bburx=544pt,bbury=540pt} 
& \hspace*{8mm}
\psfig{figure=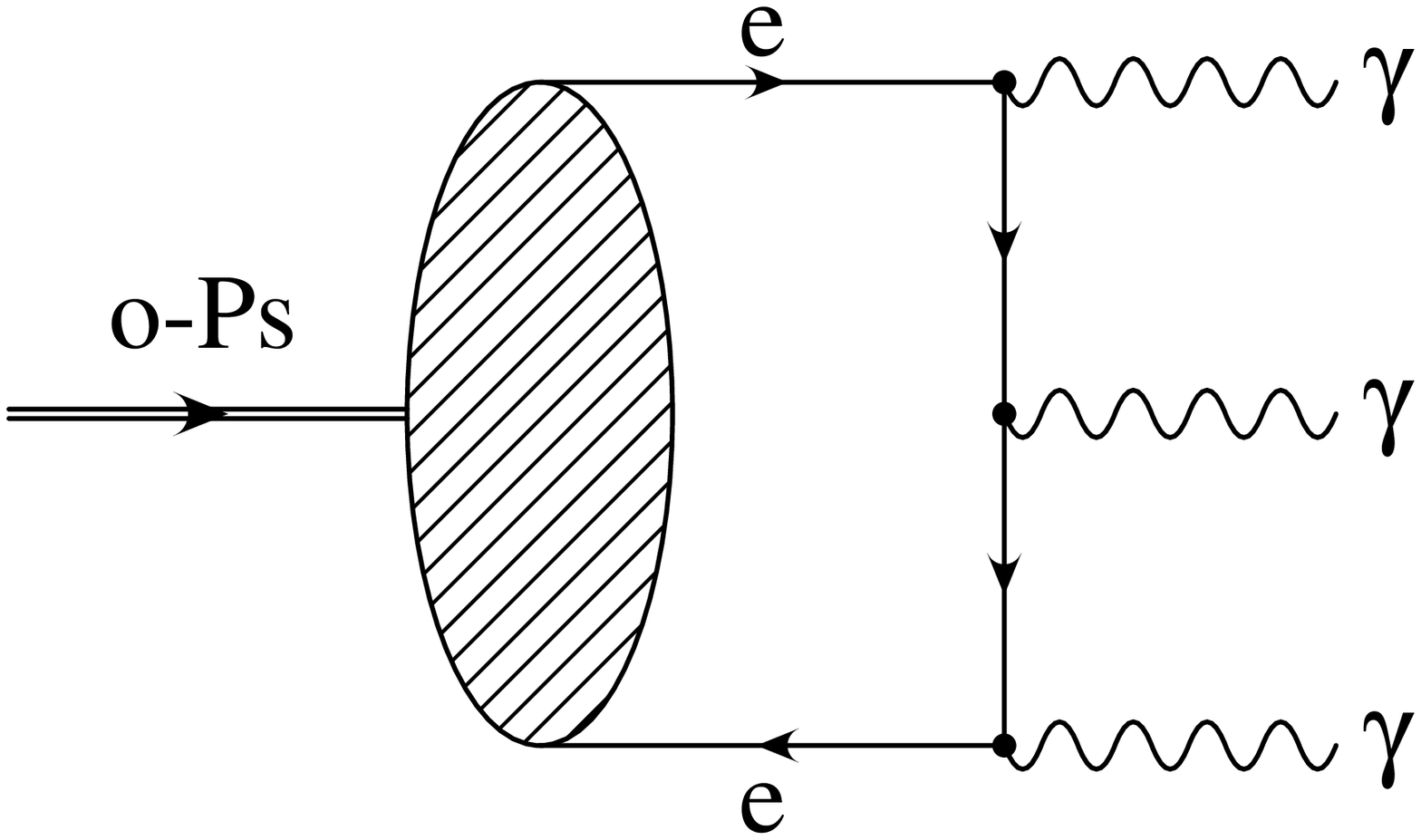,width=35mm,bbllx=72pt,bblly=291pt,%
bburx=544pt,bbury=540pt} 
\end{tabular}
\end{minipage}
\caption{Lowest order decay channels of p-Ps and o-Ps.}
\label{fig1}
\end{figure}

In the $S$ state of Ps, spins of the electron and positron can combine
to give either a total spin 0 singlet state (parapositronium, p-Ps),
or a spin 1 triplet (orthopositronium, o-Ps).  Because of the
possibility of $e^+e^-$ annihilation, both states are unstable.
Barring $C$-violating effects (e.g. caused by the weak interactions),
the ground state of parapositronium (singlet) can annihilate into only
even number of photons (see Fig.~\ref{fig1}).

The decay rate of the p-Ps ground state, $1^1S_0$, can be calculated
as a series in $\alpha$.  The two-photon decay rate is
\ba
\Gamma(\mbox{p-Ps}\to \gamma\gamma) &=&
{\alpha^5 m_e\over 2}
\left[ 1-\left(5-{\pi^2\over 4}\right) {\alpha\over \pi}
+2\alpha^2 \ln {1\over \alpha}
+ 1.75(30) \left( {\alpha\over\pi}\right)^2
- {3\alpha^3 \over 2\pi} \ln^2 {1\over \alpha}
\right.
\nonumber \\
&&
\left.
+\order{\alpha^3\ln {1\over \alpha} } \right]
\nonumber \\
&=& 7989.50(2)~\mu{\rm s}^{-1}\,,
\label{eq:pps}
\ea
where the non-logarithmic terms $\order{\alpha^2}$
\cite{Czarnecki:1999gv,Czarnecki:1999ci} and leading-logarithmic terms
$\order{\alpha^3\ln^2\alpha}$ \cite{DL} have been obtained only
recently.  
The four--photon branching ratio is of the  relative order $\alpha^2$
\cite{Billoire:1978wq,Muta:1982hb,AdkBr}: 
\begin{equation} 
\label{B24}
{\rm BR}(\mbox{p-Ps}\to 4\gamma)=
{\Gamma(\mbox{p-Ps}\to 4\gamma)
\over 
\Gamma(\mbox{p-Ps}\to \gamma\gamma)} = 
0.277(1)\left(\frac{\alpha}{\pi}\right)^2
\simeq 1.49\cdot 10^{-6}\,.
\end{equation}
The theoretical prediction, (\ref{eq:pps}), agrees well
with the experiment \cite{AlRam94},
\begin{equation}
\Gamma_{\rm exp}(\pPs)= 7990.9(1.7)~\mu{\rm s}^{-1}\,.
\end{equation}

\subsection{Orthopositronium decays}
The ground state of the orthopositronium, $1^3S_1$, 
can decay into an odd number of the photons only (if $C$ is conserved).
The three--photon (see Fig.~\ref{fig1}) decay rate formula is
\cite{Adkins96,DL} 
\ba
\Gamma(\mbox{o-Ps}\to \gamma\gamma\gamma)\!\!\!\! &=&\!\!\!\!
{2(\pi^2-9)\alpha^6 m_e\over 9\pi}
\left[ 1-10.28661(1){\alpha\over \pi}
%6\,606(10){\alpha\over \pi}
-{\alpha^2\over 3}\ln {1\over \alpha}
+ B_o \left( {\alpha\over\pi}\right)^2
- {3\alpha^3 \over 2\pi} \ln^2 {1\over \alpha}
\right.
\nonumber \\
&&
\left.
+\order{\alpha^3\ln \alpha}\right] 
\nonumber \\
&\simeq &
\left(
 7.0382  +   0.39 \cdot 10^{-4} \; B_o
\right)
\,\mu {\rm s}^{-1}\,.
\label{eq:otheor}
\ea
Because of the three-body phase space and a large number of diagrams,
a theoretical analysis of o-Ps decays is more difficult than in the
case of p-Ps.  The non-logarithmic two-loop effects, parameterized by
$B_o$, have not been evaluated yet, except for a subset of the so-called
soft corrections.  Those partial results depend on the scheme adopted
for regularizing ultraviolet divergences and do not give a reliable
estimate of the complete $B_o$.  Further theoretical work is needed to
find that potentially important correction.

\begin{table}[thb]
\begin{center}
\begin{tabular}{||l|c|l|r||}
\hline \hline
&&&\\
Reference & Method & $\Gamma(\mbox{o-Ps})~[\mu s^{-1}]$ & $B_o$ \\
&&&\\
\hline
&&&\\
London \protect{\cite{griffith78}} &Gas&7.0450(60)&174(154)\\
&&&\\
Mainz \protect{\cite{Hasbach:1987bh}}    &Vacuum&7.0310(70)&$-185(179)$\\
&&&\\
Ann Arbor \protect{\cite{Westbrook89}}&Gas&7.0514(14)&338(36)\\
&&&\\
Ann Arbor \protect{\cite{Nico:1990gi}}  &Vacuum&7.0482(16)&256(41)\\
&&&\\
Tokyo \protect{\cite{Asai:1995re}}  &Powder&7.0398(29)&41(74)\\
&&&\\
\hline
\multicolumn{2}{||c|}{}&&\\
\multicolumn{2}{||c|}{Theory without $\alpha^2$}& 7.0382&0\\
\multicolumn{2}{||c|}{}&&\\
\hline
\hline
\end{tabular}
\caption{Some experimental results for the o-Ps lifetime.  ``Method''
in the second column refers to the medium in which o-Ps decays.  The
last column shows the value of the two-loop coefficient $B_o$,
necessary to bring the theoretical prediction
(\protect\ref{eq:otheor}) into agreement with the 
given experimental value.  The last line gives the theoretical
prediction with $B_o=0$.}
\label{Tortho}
\end{center}
\end{table}

Five-photon decay branching ratio is of order $\alpha^2$
\cite{AdkBr,Lepage:1983yy}, 
\begin{equation}  \label{B35}
{\rm BR}(\mbox{o-Ps}\to 5\gamma)={\Gamma(\mbox{o-Ps}\to 5\gamma)
\over \Gamma(\mbox{o-Ps}\to  \gamma\gamma\gamma)} 
= 0.19(1)\left( {\alpha\over \pi}\right)^2
\simeq  1.0\times 10^{-6}\,,
\end{equation}
and does not significantly influence the total width.

Experimentally, o-Ps lifetime is somewhat easier to measure than that
of p-Ps, because o-Ps lives about 1000 times longer.  However, the
interactions of o-Ps with the cavity walls and/or the buffer gas are more
important.  Several experiments have been performed, but their results
are in part inconsistent with one another.  The present experimental
situation is summarized in Table~\ref{Tortho}.  
The last column indicates the value of the two-loop coefficient $B_o$
necessary to reconcile a given experimental value with the
theoretical prediction (\ref{eq:otheor}).  We see that the most
precise Ann Arbor experiments require an anomalously large value of
$B_o$.  This has been known as the ``o-Ps lifetime puzzle.''

It should be mentioned that the perturbative coefficients are moderate
in all known QED predictions for observables studied with high
accuracy (e.g. anomalous magnetic moments of the electron and muon,
Lamb shift in the hydrogen atom, or the ground state hyperfine
interval in muonium). The known corrections to the positronium gross,
fine \cite{PhK,Czarnecki:1999mw} and hyperfine structure
\cite{Adkins97,Hoang:1997ki,PhK,Czarnecki:1998zv} and to
parapositronium decay rate 
\cite{Czarnecki:1999gv,Czarnecki:1999ci} are also moderate.  In any
case, before definite conclusions about comparison of theory and
experiment can be made, it is necessary to clarify the experimental
situation.  Efforts to improve the experiments are underway
\cite{AsaiPriv,skalsey98}.

\section{Exotic decays of positronium}
The o-Ps lifetime found in the precise experiments in Ann Arbor 
\cite{Westbrook89,Nico:1990gi} is significantly shorter than
expected in QED.  It has been speculated that there might be
other decay channels which contribute to the faster decay rate.  A
number of dedicated experimental searches of rare Ps decays were
undertaken. 

An example of rare decays predicted by QED are the multi-photon
channels.  For p-Ps the experimental results are
\ba
{\rm BR}(\mbox{p-Ps}\to 4\gamma)
&\simeq & 1.48(18)\cdot 10^{-6}\,, \qquad  \cite{Adachi:1990nj,Adachi94,Chib},
\nonumber \\
{\rm BR}(\mbox{p-Ps}\to 4\gamma)
&\simeq & 1.50(11)\cdot 10^{-6}\,,  \qquad  \cite{Busc},
\ea
in agreement with the theoretical expectation (\ref{B24}). For o-Ps
one also finds agreement, with (\ref{B35}):
\begin{equation}
{\rm BR}(\mbox{o-Ps}\to 5\gamma)
\simeq 2.2^{+2.6}_{-1.8}\cdot 10^{-6}\,, \qquad \cite{Mats,Chib}\,.
\end{equation}

There have also been a number of searches for forbidden photonic
decays and for modes with some unregistered particles.  Some of such
studies for o-Ps are summarized in Tables \ref{Texo} and \ref{Tless}.
In Table \ref{Texo} we list searches for decay modes involving an
exotic spinless particle $X$, accompanied by a single photon.  
Table \ref{Tless} refers to decays into invisible particles.

Standard Model predicts some ``invisible'' decay channels, namely a
conversion of Ps into a neutrino--antineutrino pair.    We present a
calculation of the corresponding decay rate in the next section.

\begin{table}[t]
\begin{center}
\begin{tabular}{||c|c|r|r||}
\hline \hline
&&&\\[-1ex]
Ref. & Mode & Branching ratio& Boson mass   \\[1ex]
\hline
&&&\\[-1ex]
\protect{\cite{Orito:1989pc}} &$X^L+\gamma$
  &$<6.4 \cdot10^{-5} - 7.6 \cdot10^{-6}$ (at 90\% CL) & $m_X<800$ keV\\[1ex]
\protect{\cite{Tsuchiaki:1990bw}} &$X+\gamma$
  &$<\Big(1\!-\!6\Big)\cdot10^{-4}$ (at 95\% CL)&$300<m_X<900$ keV\\[1ex]
\protect{\cite{Asai:1991rd}} &$X^L+\gamma$
  &$<1.1\cdot10^{-6}$ (at 90\% CL) &$m_X<800$ keV\\[1ex]
\protect{\cite{Asai:1994nw}} &$X^S+\gamma$
  &$<3.0\cdot10^{-4}$ (at 90\% CL) &$m_X<500$ keV\\[1ex]
\protect{\cite{Maeno:1995gf}} &$X^S+\gamma$
  &$<2.0\cdot10^{-4}$ (at 90\% CL)&$847<m_X< 1013$ keV\\[1ex]
\hline \hline
\end{tabular}
\end{center}
\caption{\label{Texo} Searches for exotic decays 
of orthopositronium with one
detected photon. $X$ denotes a neutral boson: $L$ --
long-lived and $S$ -- short-lived. }
\end{table}

\begin{table}[htb]
\begin{center}
\begin{tabular}{||c|c|c||}
\hline \hline
&&\\[-1ex]
Ref. & Mode & Branching ratio  \\[1ex]
\hline
&&\\[-1ex]
\protect{\cite{Ato}} &$\mbox{o-Ps}\to X$
  &$<5.8\cdot10^{-4}$ (at 90\% CL)\\[1ex]
\protect{\cite{Mits93}} &$\mbox{p-Ps}\to X$
  &$<1.7\cdot10^{-6}$ (at 90\% CL)\\[1ex]
\protect{\cite{Mits93}} &$\mbox{o-Ps}\to X$
  &$<2.8\cdot10^{-6}$ (at 90\% CL)\\[1ex]
\hline \hline
\end{tabular}
\end{center}
\caption{\label{Tless} Searches for invisible decays of Ps. $X$
denotes any invisible products.}
\end{table}

\section{Weak decays of orthopositronium}

Of the two spin states of Ps, only o-Ps (spin 1 state) 
can decay weakly into a pair
$\nu\bar\nu$, at least if the neutrinos are massless and if there is no
emission of photons.

Two amplitudes contribute to the decay  $\oPs\to \nu_e\bar \nu_e$: $W$
exchange in $t$ channel and an annihilation via $Z$.
The rate of this decay is
\begin{equation}
\Gamma(\oPs\to \nu_e\bar \nu_e) = {1\over 3M^2} {1\over 8\pi}
|\psi(0)|^2 |{\cal M}|^2\,,
\end{equation}
where the factor $1/3$ is due to the average over the spin orientations,
$M\approx 2m_e$ is the mass of the Ps, $1/8\pi$ comes from the phase
space, and the square of the wave function at the origin is
\begin{equation}
|\psi(0)|^2 = {1\over \pi a^3} = {\alpha^3 m_e^3\over 8\pi}\,,
\end{equation}
where $a={2\over\alpha m_e}$ is the Bohr radius of the Ps.
The transition matrix element is given by
\begin{eqnarray}
{\cal M} ={iG_F\over \sqrt{2}}
\left[
\bar u_\nu \gamma^\mu (1-\gamma_5) u_e    \right. \!\!\!\! &\cdot &
\!\!\!\!
\bar v_e \gamma_\mu (1-\gamma_5) v_\nu
\nonumber\\
 \qquad
-
\bar u_\nu \gamma^\mu (1-\gamma_5) v_\nu \!\!\!\! &\cdot & \!\!\!\!
\left.
\bar v_e \gamma_\mu
  \left( 2s^2-{1\over 2} + {1\over 2}\gamma_5 \right)
u_e
\right]\,,
\\
s\equiv \sin\theta_W\,.\qquad\qquad \qquad &&
\end{eqnarray}

The minus sign between the two terms is due to the different ordering
of the fermion fields.  It leads to a destructive interference of
the $W$ and $Z$ exchange amplitudes.  The decay rate is
\begin{equation}
\Gamma(\oPs\to \nu_e\bar \nu_e) = {G_F^2\alpha^3 m_e^5\over 24\pi^2}
\left(1+4s^2\right)^2\,,
\label{eq:ee}
\end{equation}
which corresponds to a totally negligible branching ratio of about
$6.2\times 10^{-18}$. 
We note that after Fierz transformation of the $W$ exchange diagram,
the effective Hamiltonian for the $\oPs$ decay coincides with the well
known Hamiltonian for the neutrino-electron scattering \cite{Cheng}
\begin{eqnarray}
{\cal H}_{\eff}  &=& {G_F\over \sqrt{2}}
\bar u_\nu \gamma^\mu (1-\gamma_5) v_\nu \cdot
\bar v_e \gamma_\mu
(a - b\gamma_5) u_e\,,
\\
a&=& 2s^2+{1\over 2}\,, \qquad b={1\over 2}\,.
\end{eqnarray}

The decay $\oPs\to \nu_e\bar \nu_e$ 
was discussed in  Ref.~\cite{Govaerts,PisaProblems}.  Our
result differs from both publications.  In Ref.~\cite{Govaerts} the
relative minus sign between the $W$ and $Z$ exchange amplitudes was
left out, so that the interference was constructive.  In Ref. 
\cite{PisaProblems} there is a mistake in the trace calculation so
that the decay seems to arise only due to the axial, rather than
vector, part of the effective Hamiltonian.

For other flavors of the final state neutrinos only the annihilation
diagram contributes.  Since the vector coupling of $Z$ to electrons is
proportional to $1-4s^2$, these channels are strongly suppressed
compared to the electron--neutrino final state.  We find for $l\neq e$
\begin{equation}
\Gamma(\oPs\to \nu_l \bar \nu_l) = {G_F^2\alpha^3 m_e^5\over 24\pi^2}
\left(1-4s^2\right)^2.
\label{eq:othernu}
\end{equation}
The branching ratio for each non-electron neutrino flavor is about
$9.5\times 10^{-21}$, much smaller than the experimental limits 
(cf. Table \ref{Tless}).  Certainly it cannot
explain the ``orthopositronium lifetime puzzle.''

\section{Weak decays of other leptonic bound states}

In this section we briefly summarize the influence of weak
interactions on the lifetimes of the $\mu^+\mu^-$ and $\mu^+e^-$
bound states.

Dimuonium (Dm) is a bound state of a muon and an anti-muon. The
spectrum \cite{jentschura97,Karshenboim:1998we}, QED
\cite{Karshenboim:1998we,Ginzburg:1998df} decay modes and possible
production \cite{Ginzburg:1998df} of such a system have been studied
recently.  In contrast to orthopositronium, the decay rate of
orthodimuonium ($\oDm$) is of order $\alpha^5m_\mu$ because of
conversion into the the electron--positron pair via a virtual
single--photon annihilation,
\begin{equation}
\Gamma^{(0)}(\oDm\to e^+e^-)  = \frac{ \alpha^5 m_\mu }{6} \,.
\end{equation}
In addition to the dominant QED decay channel, $\oDm$ can
decay into neutrinos or electrons via an annihilation amplitude with a
$Z$-boson exchange.    The decay width into muon
neutrinos is given by a formula analogous to eq.~(\ref{eq:ee}), with
muon mass substituted for $m_e$; the width into other neutrino flavors
is obtained similarly from eq.~(\ref{eq:othernu}).

In case of the decay o-Dm$\to e^+e^-$, the $Z$ contribution is a
tiny correction to the photon exchange:
\[
\Gamma(\oDm\to e^+e^-)
={ \alpha^5m_\mu  \over 6}\left(
1+{G_Fm_\mu^2(1-4s^2)^2\over 4\sqrt{2}\pi\alpha} 
+{G_F^2m_\mu^4(1-4s^2)^2\left[1+(1-4s^2)^2\right]\over 32\pi^2\alpha^2}
\right).
\]
The two correction terms correspond, respectively, to the interference
between the photon and $Z$ amplitudes, and to the square of the $Z$
amplitude.

Muonium (Mu) consists of an electron and an antimuon.  Its decays have
recently been discussed in some detail in Ref.~\cite{Czarnecki:1999yj}.
The electroweak interactions can lead to its decay
via a $W$ exchange in the $t$ channel.  For the  decay width of a
bound state of particles with unequal masses we use
\begin{equation}
\Gamma(\oMu\to \nu_e\bar \nu_\mu) = {1\over 3\cdot 4m_\mu m_e}
{1\over 8\pi} |\psi(0)|^2 |{\cal M}|^2
\end{equation}
which leads to
\begin{equation}
\Gamma(\oMu\to \nu_e\bar \nu_\mu) = {G_F^2\alpha^3 m_e^3m_\mu^3\over
 3\pi^2(m_\mu+m_e)}\,.
\end{equation}
This formula can also be derived from known results in $B_c$ decays
\cite{Beneke:1996xe}.  

\section{Summary}
We have reviewed the confrontation of theoretical and experimental
results concerning decays of positronium.  We believe that the
``orthopositronium lifetime puzzle'' is not likely to be solved by
large two-loop QED corrections, nor by a discovery of an exotic decay
channel.  We have investigated in some detail the annihilation
amplitude into neutrinos and found that its contribution is
negligible.  In fact it is even smaller than the results previously
cited in the literature.  Most likely, the situation will be clarified
by the future improved experiments, and we are looking forward to
their results.

\section*{Acknowledgements}
A.C. thanks S. Davidson for carefully reading the manuscript and
useful remarks, and K. Melnikov and A. Yelkhovsky for collaboration on
positronium physics. S. K. is grateful to D. Gidley for stimulating
discussions.  The work of A.C. was supported by the DOE grant
DE-AC02-98CH10886, and the work of S.K. was supported in part by the
Russian State Program `Fundamental Metrology' and NATO grant CRG
960003.

%\bibliographystyle{../../pro/tex/revtex}
%\bibliography{../../pro/tex/phd}

\end{document}